\documentclass[preprint,pra,showpacs,showkeys]{revtex4}
\usepackage{graphicx,amsmath,amsfonts,amssymb}
\usepackage{times}
\begin{document}

\title{The controlled teleportation of an arbitrary two-atom  entangled state in driven cavity QED}
\author{Chuan-Jia Shan\footnote{E-mail: shanchuanjia1122@yahoo.com.cn}, Ji-Bing Liu, Tang-Kun Liu, Yan-Xia Huang, Hong Li}
\affiliation{College of Physics and Electronic Science, Hubei Normal
University, Huangshi 435002, China}
\date{\today}

\begin{abstract}
In this paper, we propose a scheme for the controlled teleportation
of an arbitrary two-atom entangled state
$|\phi\rangle_{12}=a|gg\rangle_{12}+b|ge\rangle
_{12}+c|eg\rangle_{12}+d|ee\rangle_{12}$ in driven cavity QED.  An
arbitrary two-atom  entangled state can  be teleported perfectly
with the help of the cooperation of the third
 side  by constructing a three-atom GHZ entangled state as the controlled channel.
 This scheme does
not involve apparent (or direct) Bell-state measurement and is
insensitive to the cavity decay and the thermal field. The
probability of the success in our scheme is 1.0.
\end{abstract}



\maketitle

\section{Introduction}

\indent Since the first scheme for quantum teleportation was
proposed by Bennett et al.[1], all kinds of schemes[2-5] have been
proposed and experimental demonstration of quantum teleportation has
been realized with the polarization photon[6] and a single coherent
mode of a field in optical systems[7] and NMR[8]. In 1998, Karlsson
and Bourennane[9] showed that an arbitrary unknown state of a qubit
could be teleported to either one of two receivers by the use of a
three-qubit entangled Greenberger-Horne-Zeilinger (GHZ) state. One
of the two agents acts as the controller and the other recovers the
unknown state according to the information published by the sender
and the controller. Since that work, many investigations have been
made on the controlled teleportation[10-12].

Recently, Riebe et al.[13] and Barrett et al.[14] have implemented
the first experimental realization of the teleportation of atomic
qubits in ion-trap system, which will attract more attention for
quantum information processing on the field of cavity QED. In cavity
QED, schemes have been proposed for teleportation of two-particle
entangled states[15] and multipartite entangled atomic states[16].
However, the main experimental challenge consists in the so-called
Bell-state measurement. Zheng[17] and Ye[18] proposed a scheme for
teleporting an unknown atomic state in cavity QED without Bell-state
measurement, but the probability of success is only 0.25 and 0.5.
Jin et al.[19] made a proposal for teleporting two-atom entangled
state with a probability 1.0 by adding a classical driving field.

As for the arbitrary two-atom two-level entangled state, Lee et
al.[20] showed that it was possible to teleport a two-qubit state
$|\phi\rangle
_{12}=a|gg\rangle_{12}+b|ge\rangle_{12}+c|eg\rangle_{12}+d|ee\rangle_{12}$
from Alice to Bob using a four-entangled state and sending to him
four bits of classical information. Rigolin[21] explicitly
constructs this protocol and presents a generalization to N qubits,
but this protocol needs a set of 16 generalized Bell states to
implement the teleportation. Deng[22] present a way for symmetric
multiparty-controlled teleportation of an arbitrary two-particle
entangled state based on Bell-basis measurements by using two
Greenberger-Horne-Zeilinger states. In this paper, we propose a
scheme for the controlled teleportation of an arbitrary two-atom
two-level entangled state in driven cavity QED. In contrast to the
previous scheme, the present one has the following advantages: First
such an arbitrary and unknown two-atom state is transmitted from a
sender to a receiver, whereas the scheme in Ref.[19] considers a
special unknown two-atom state. Second, the quantum channel is
different, in our protocol, the quantum channel is composed of the
 two-atom maximally entangled state comparing with the previous
scheme[21] using a four-entangled state. Whereas two-atom maximally
entangled state is much easier to prepare and maintain than
four-atom entangled state. Finally the  protocol does not involve
apparent (or direct) Bell-state measurement and is insensitive to
the cavity decay and the thermal field. The probability of the
success can reach 1.0.
\section{the model}
Consider two two-level atoms interacting resonantly with a
single-mode cavity field, at the same time, the two atoms are driven
by a classical field. The interaction between atoms and the cavity
can be described as follows[23]:
\begin{equation}
H=\omega_{0}\sum_{j=1}^{2}S_{j}^{z}+\omega_{a}a^{\dag}a+\sum_{j=1}%
^{2}[g(a^{\dag}S_{j}^{-}+aS_{j}^{+})+\Omega(S_{j}^{+}e^{-i\omega_{d}t}%
+S_{j}^{-}e^{i\omega_{d}t})].
\end{equation}
where $\omega_{0}$, $\omega_{a}$ and $\omega_{d}$ are atomic
transition frequency, cavity frequency and the frequency of driving
field, respectively, $a^{\dag}$ and $a$ are creation and
annihilation operators for the cavity mode, $g$ is the coupling
constant between atoms and cavity, atomic operators
$S_{j}^{+}=|e\rangle_{j}\langle g|$,$S_{j}^{-}=|g\rangle_{j}\langle e|$%
,$S_{j}^{z}=\frac{1}{2}(|e\rangle_{j}\langle e|-|g\rangle_{j}\langle
g|)$ , and $\Omega$ is the Rabi frequency of the classical field. We
consider the case $\omega_{0}=\omega_{d}$. In the interaction
picture, the evolution operator of the system is[23]
\begin{equation}
U(t)=e^{-iH_{0}t}e^{-iH_{e}t}.
\end{equation}
where $H_{0}=\sum_{j=1}^{2}\Omega(S_{j}^{-}+S_{j}^{+})$, $H_{e}$ is
the effective Hamiltonian. In the large detuning
$\delta\gg\frac{1}{2}g$ and strong driving field
$2\Omega\gg\delta,g$ limit, the effective Hamiltonian for this
interaction can be described as follows[23]:
\begin{equation}
H_{e}=\frac{1}{2}\lambda\lbrack\sum_{j=1}^{2}(|e\rangle_{j}\langle
e+|g\rangle_{j}\langle g)+\sum_{j,k=1,j\neq k}^{2}(S_{j}^{+}S_{k}^{+}%
+S_{j}^{+}S_{k}^{-})+H.c].
\end{equation}
where $\lambda=g^{2}/2\delta$, $\delta$ is the detuning between
$\omega_{0}$ and $\omega_{a}$. From Eq.(3), we know that $H_{e}$ is
independent of creation and annihilation operators of the cavity
mode and is only related with atomic operators. So the effects of
cavity decay and thermal field are all eliminated. When two atoms
exist in the cavity, after interaction time t, the state of the two
atoms will undergo the following evolution:
\begin{eqnarray}
|ee\rangle_{jk}  &  \longrightarrow e^{-i\lambda t}[\cos\lambda
t(\cos\Omega t|e\rangle_{j}-i\sin\Omega
t|g\rangle_{j})\times(\cos\Omega t|e\rangle
_{k}-i\sin\Omega t|g\rangle_{k})\nonumber\\
&  -i\sin\lambda t(\cos\Omega t|g\rangle_{j}-i\sin\Omega t|e\rangle_{j}%
)\times(\cos\Omega t|g\rangle_{k}-i\sin\Omega t|e\rangle_{k})], \label{8}%
\end{eqnarray}%
\begin{eqnarray}
|eg\rangle_{jk}  &  \longrightarrow e^{-i\lambda t}[\cos\lambda
t(\cos\Omega t|e\rangle_{j}-i\sin\Omega
t|g\rangle_{j})\times(\cos\Omega t|g\rangle
_{k}-i\sin\Omega t|e\rangle_{k})\nonumber\\
&  -i\sin\lambda t(\cos\Omega t|g\rangle_{j}-i\sin\Omega t|e\rangle_{j}%
)\times(\cos\Omega t|e\rangle_{k}-i\sin\Omega t|g\rangle_{k})], \label{9}%
\end{eqnarray}%
\begin{eqnarray}
|ge\rangle_{jk}  &  \longrightarrow e^{-i\lambda t}[\cos\lambda
t(\cos\Omega t|g\rangle_{j}-i\sin\Omega
t|e\rangle_{j})\times(\cos\Omega t|e\rangle
_{k}-i\sin\Omega t|g\rangle_{k})\nonumber\\
&  -i\sin\lambda t(\cos\Omega t|e\rangle_{j}-i\sin\Omega t|g\rangle_{j}%
)\times(\cos\Omega t|g\rangle_{k}-i\sin\Omega t|e\rangle_{k})], \label{10}%
\end{eqnarray}%
\begin{eqnarray}
|gg\rangle_{jk}  &  \longrightarrow e^{-i\lambda t}[\cos\lambda
t(\cos\Omega t|g\rangle_{j}-i\sin\Omega
t|e\rangle_{j})\times(\cos\Omega t|g\rangle
_{k}-i\sin\Omega t|e\rangle_{k})\nonumber\\
&  -i\sin\lambda t(\cos\Omega t|e\rangle_{j}-i\sin\Omega t|g\rangle_{j}%
)\times(\cos\Omega t|e\rangle_{k}-i\sin\Omega t|g\rangle_{k})].
\end{eqnarray}

\section{The controlled teleportation of an arbitrary two-atom entangled state}
Assume that the two atoms to be teleported are initially in the
state
\begin{equation}
|\phi \rangle _{12}=a|gg\rangle _{12}+b|ge\rangle _{12}+c|eg\rangle
_{12}+d|ee\rangle _{12}
\end{equation}%
where $a$, $b$, $c$ and $d$ are unknown coefficients, $\left\vert
a\right\vert ^{2}+\left\vert b\right\vert ^{2}+\left\vert
c\right\vert ^{2}+\left\vert d\right\vert ^{2}=1$. $|e\rangle $ and
$|g\rangle $ are the excited and ground states of the atom.  The
quantum channels are
 a  three-atom maximally entangled state and a  two-atom maximally entangled state.
\begin{equation}
|\phi\rangle_{345}=\frac{1}{\sqrt{2}}(|ggg\rangle_{345}+i|eee\rangle_{345}),
\end{equation}
\begin{equation}
|\phi\rangle_{67}=\frac{1}{\sqrt{2}}(|ge\rangle_{67}-i|eg\rangle_{67}).
\end{equation}
Here the atoms 1, 2, 3 and 6 belong to the sender Alice, atoms 4, 7
belong to the receiver Bob, and atom 5 the controlling atom belongs
to Charlie. The initial state of the whole system is given by
\begin{eqnarray}
|\psi\rangle_{1234567}&=\frac{1}{2}(a|gg\rangle_{12}+b|ge\rangle
_{12}+c|eg\rangle_{12}+d|ee\rangle_{12})\otimes(|ggg\rangle_{345}
+i|eee\rangle_{345})\nonumber\\
&\otimes(|ge\rangle_{67}-i|eg\rangle_{67}).
\end{eqnarray}

Then Alice sends atoms 1, 3 into a single-mode cavity and atoms 2, 6
into another cavity. We only write the time evolution of the first
term part$|gggggge\rangle_{1234567}$ of Eq.(11).
\begin{eqnarray}
|gggggge\rangle_{1234567}  &  =e^{-2i\lambda t}[\cos\lambda
t(\cos\Omega t|g\rangle_{1}-i\sin\Omega t|e\rangle_{1})(\cos\Omega
t|g\rangle
_{3}-\nonumber\\
&  i\sin\Omega t|e\rangle_{3})-i\sin\lambda t(\cos\Omega t|e\rangle_{1}%
-i\sin\Omega t|g\rangle_{1})\times\nonumber\\
&  (\cos\Omega t|e\rangle_{3}-i\sin\Omega
t|g\rangle_{3})]\times\lbrack
\cos\lambda t(\cos\Omega t|g\rangle_{2}-\nonumber\\
&  i\sin\Omega t|e\rangle_{2})(\cos\Omega t|g\rangle_{6}-i\sin\Omega
t|e\rangle_{6})-i\sin\lambda t\times\nonumber\\
&  (\cos\Omega t|e\rangle_{2}-i\sin\Omega t|g\rangle_{2})(\cos\Omega
t|e\rangle_{6}-i\sin\Omega t|g\rangle_{6})]\times\nonumber\\
&  |gge\rangle_{457}.
\end{eqnarray}
In the same method, we can choose $\lambda t=\frac{1}{4}\pi$,
$\Omega t=\pi$ by modulating the driving field appropriately. The
whole system will evolve into
\begin{eqnarray}
& |\psi\rangle=\frac{1}{{4}}[|eeee\rangle_{1326}\times(-a|gge\rangle
_{457}-b|ggg\rangle_{457}+c|eee\rangle_{457}+d|eeg\rangle_{457})+\nonumber\\
& |eegg\rangle_{1326}\times(-ia|gge\rangle_{457}+i b|ggg\rangle_{457}%
+ic|eee\rangle_{457}-id|eeg\rangle_{457})+\nonumber\\
&  |ggee\rangle_{1326}\times(-ia|gge\rangle_{457}-i b|ggg\rangle_{457}%
-ic|eee\rangle_{457}-id|eeg\rangle_{457})+\nonumber\\
&  |gggg\rangle_{1326}\times(a|gge\rangle_{457}-b|ggg\rangle_{457}%
+c|eee\rangle_{457}-d|eeg\rangle_{457})+  \nonumber\\
&  |egeg\rangle_{1326}\times(-a|eeg\rangle_{457}+b|eee\rangle_{457}%
-c|ggg\rangle_{457}+d|gge\rangle_{457})+\nonumber\\
&  |egge\rangle_{1326}\times(-ia|eeg\rangle_{457}-i b|eee\rangle_{457}%
-ic|ggg\rangle_{457}-id|gge\rangle_{457})+\nonumber\\
&  |geeg\rangle_{1326}\times(-ia|eeg\rangle_{457}+i b|eee\rangle_{457}%
-ic|ggg\rangle_{457}+id|gge\rangle_{457})+\nonumber\\
&  |gege\rangle_{1326}\times(a|eeg\rangle_{457}+b|eee\rangle_{457}%
-c|ggg\rangle_{457}-d|gge\rangle_{457})+\nonumber\\
&  |eeeg\rangle_{1326}\times(ia|ggg\rangle_{457}-i b|gge\rangle_{457}%
-ic|eeg\rangle_{457}+id|eee\rangle_{457})+\nonumber\\
&  |eege\rangle_{1326}\times(-a|ggg\rangle_{457}-b|gge\rangle_{457}%
+c|eeg\rangle_{457}+d|eee\rangle_{457})+\nonumber\\
&  |ggeg\rangle_{1326}\times(-a|ggg\rangle_{457}+b|gge\rangle_{457}%
-c|eeg\rangle_{457}+d|eee\rangle_{457})+\nonumber\\
&  |ggge\rangle_{1326}\times(-ia|ggg\rangle_{457}-i b|gge\rangle_{457}%
-ic|eeg\rangle_{457}-id|eee\rangle_{457})+\nonumber\\
&  |egee\rangle_{1326}\times(-ia|eee\rangle_{457}-i b|eeg\rangle_{457}%
-ic|gge\rangle_{457}-id|ggg\rangle_{457})+\nonumber\\
&  |eggg\rangle_{1326}\times(a|eee\rangle_{457}-b|eeg\rangle_{457}%
+c|gge\rangle_{457}-d|ggg\rangle_{457})+\nonumber\\
&  |geee\rangle_{1326}\times(a|eee\rangle_{457}+b|eeg\rangle_{457}%
-c|gge\rangle_{457}-d|ggg\rangle_{457})+\nonumber\\
&  |gegg\rangle_{1326}\times(ia|eee\rangle_{457}-i b|eeg\rangle_{457}%
-ic|gge\rangle_{457}+id|ggg\rangle_{457})].
\end{eqnarray}

In order to realize the teleportation, Alice should make a separate
measurement on atoms 1, 2, 3 and 6. If Alice detects the atoms in
the state $|eeee\rangle_{1326}$, after measurement the state of
atoms 4, 5 and 7 will collapse into
\begin{equation}
-a|gge\rangle_{457}-b|ggg\rangle_{457}+c|eee\rangle_{457}+d|eeg\rangle_{457}.
\end{equation}

Now Alice informs Bob and Charlie of the result of the measurement
by the classical channels, after Charlie receives the information,
if Charlie would like to help Bob with the teleportation, he should
perform the Hadamard operation in the following forms on atom 5 in
the basis ${|g\rangle,|e\rangle}$.
\begin{equation}
H|g\rangle=\frac{1}{\sqrt{2}}(|g\rangle+|e\rangle),H|e\rangle=\frac{1}
{\sqrt{2}}(|g\rangle-|e\rangle),
\end{equation}
Eq.(14) will become
\begin{eqnarray}
&  -\frac{1}{\sqrt{2}}[|g\rangle_{5}(a|ge\rangle_{47}+b|gg\rangle
_{47}-c|ee\rangle_{47}-d|eg\rangle_{47})+\nonumber\\
&  |e\rangle_{5}(a|ge\rangle_{47}+b|gg\rangle_{47}+c|ee\rangle_{47}%
+d|eg\rangle_{47})].
\end{eqnarray}

If Charlie measures atom 5 in the $|e\rangle_{5}$, the atom 4 and 7
are in the state
\begin{equation}
a|ge\rangle_{47}+b|gg\rangle_{47}+c|ee\rangle_{47}+d|eg\rangle_{47},
\end{equation}

Then Charlie tells the measurement result to Bob via classical
channel, Bob makes a unitary transformation
$I_{4}\otimes\sigma_{7x}$ on atoms 4, 7 to recover the teleported
state, Eq.(17) will become
\begin{equation}
a|gg\rangle_{47}+b|ge\rangle_{47}+c|eg\rangle_{47}+d|ee\rangle_{47}.
\end{equation}
If Charlie measures atom 5 in the $|g\rangle_{5}$, the atoms 4 and 7
are in the state
\begin{equation}
a|ge\rangle_{47}+b|gg\rangle_{47}-c|ee\rangle_{47}-d|eg\rangle_{47},
\end{equation}
Then Bob makes a unitary transformation
$\sigma_{4z}\otimes\sigma_{7x}$ on atoms 4, 7, Eq.(19) will become
\begin{equation}
a|gg\rangle_{47}+b|ge\rangle_{47}+c|eg\rangle_{47}+d|ee\rangle_{47}.
\end{equation}
Now the teleportation is successful. From Eq.(13) to Eq.(14), we can
see that after Alice makes a separate measurement, quantum
information of the teleported state is encoded into the state of the
atoms 4, 5, 7, which is shared between Bob and Charlie, so Bob
cannot fully recover the origin state. However, the receiver can
successfully get access to the original state if Charlie
collaborates through the local operation and classical communication
with the receiver. From Eq.(13), we can see that when atoms 1, 3 and
atoms 2, 6 enter into the cavity, because every atom has two levels,
16 kinds of different separate states can be derived. It is evident
that Bob must operate relevant unitary transformation against
Alice's and Charlie's different measurement results. By the similar
method, if the measurement results of Alice are
$|eegg\rangle_{1326}$, $|ggee\rangle_{1326}$, $|gggg\rangle_{1326}$,
$|egeg\rangle_{1326}$, $|egge\rangle_{1326}$, $|geeg\rangle_{1326}$,
$|gege\rangle_{1326}$, $|eeeg\rangle_{1326}$, $|eege\rangle_{1326}$,
$|ggeg\rangle_{1326}$, $|ggge\rangle_{1326}$, $|eggg\rangle_{1326}$,
$|egee\rangle_{1326}$, $|geee\rangle_{1326}$ and
$|gegg\rangle_{1326}$, the teleportation can also succeed with the
help of Charlie. In principle for each measurement outcome of Alice,
we should write down what measurement Charlie should do in order to
assist Alice and Bob. For the sake of saving the space, the other
outcomes and the unitary transformations are depicted in Table I.

Now we calculate the probability of successful teleportation in this
scheme. From Eq.(13) to Eq.(14), the probability of detecting the
state $|eeee\rangle _{1326}$ is 1/16, from Eq.(14) to Eq.(20), the
probability of successful teleportation is 1.0. So we can know that
the probability of successful teleportation is 1/16. The
probabilities of successful teleportation in the other fifteen
measurement results are easily derived. That is to say, the total
success probability is $1/16\times16=1.0$.

\indent Let us provide thorough physical analysis and discussions of
this scheme. Firstly let the atoms with Alice interact in a driven
cavity QED. The separate state of the two atoms may evolve into a
 two-atom maximally entangled state by properly choosing the time and
coupling constants. The separate measurements on the atoms in a
driven cavity QED substitute apparent (or direct) Bell measurements.
Therefore the difficult apparent (or direct) Bell state measurements
that Alice should perform in order to teleport his qubits are not
needed. Secondly after Charlie performs the Hadamard operation on
her atom, he can obtain two outcomes, if Charlie tells the
measurement result to Bob via classical channel, Bob only makes an
unitary transformation to recover the teleported state based on
Charlie's measurement result.

From the above analysis, we can see that Alice only needs to make a
separate measurement, therefore the difficult apparent (or
direct)Bell state measurements that Alice should perform in order to
teleport her qubits are not needed. The two qubit state can be
perfectly teleported with the help of Charlie's Hadamard operation.
That is to say, after Charlie receives this information, if Charlie
would like to help Bob with the teleportation, he should make an
operation on atom 5. That is the controlled teleportation of an
arbitrary two-atom state. All the above we suppose a three-qubit GHZ
state as the quantum channel, only a atom is supposed as the
controlling agent. In the same method, if we generalized a
(n+2)-qubit GHZ state, two qubits of (n+2)-qubit GHZ state belong to
Alice and Bob, respectively, while the other n GHZ qubits belong to
n agents, we can successfully get the arbitrary two-atom two-level
entangled state as long as all the agents cooperate and send
classical communication. Therefore the controlled teleportation is
useful in networked quantum information processing.

TABLE I. The results and the unitary transformations to finish the
controlled teleportation of an arbitrary two-atom entangled state.
$I$ is the
identity operator and $\sigma_{x,y,z}$ are the usual Pauli matrices. $U_{i}%
=|g\rangle_{i}\langle e|-|e\rangle_{i}\langle g|$ \newline%
\begin{tabular}
[c]{|c|c|c|c|c|c|}\hline Alice & Charlie & Bob operation & Alice &
Charlie  & Bob operation\\\hline $|eeee\rangle_{1326}$ &
$|e\rangle_{5}$ & $I_{4}\otimes\sigma_{7x}$ &
$|eeeg\rangle_{1326}$ & $|e\rangle_{5}$ & $I_{4}\otimes\sigma_{7z}$\\
$|eeee\rangle_{1326}$ & $|g\rangle_{5}$ &
$\sigma_{4z}\otimes\sigma_{7x}$ &
$|eeeg\rangle_{1326}$ & $|g\rangle_{5}$ & $\sigma_{4z}\otimes\sigma_{7z}$\\
$|egee\rangle_{1326}$ & $|e\rangle_{5}$ & $U_{4}\otimes\sigma_{7x}$
&
$|egeg\rangle_{1326}$ & $|e\rangle_{5}$ & $U_{4}\otimes\sigma_{7z}$\\
$|egee\rangle_{1326}$ & $|g\rangle_{5}$ &
$\sigma_{4x}\otimes\sigma_{7x}$ &
$|egeg\rangle_{1326}$ & $|g\rangle_{5}$ & $\sigma_{4x}\otimes\sigma_{7z}$\\
$|eegg\rangle_{1326}$ & $|e\rangle_{5}$ & $I_{4}\otimes U_{7}$ &
$|eege\rangle_{1326}$ & $|e\rangle_{5}$ & $I_{4}\otimes I_{7}$\\
$|eegg\rangle_{1326}$ & $|g\rangle_{5}$ & $\sigma_{4z}\otimes U_{7}$
&
$|eege\rangle_{1326}$ & $|g\rangle_{5}$ & $\sigma_{4z}\otimes I_{7}$\\
$|geee\rangle_{1326}$ & $|e\rangle_{5}$ &
$\sigma_{4x}\otimes\sigma_{7x}$ &
$|egge\rangle_{1326}$ & $|e\rangle_{5}$ & $U_{4}\otimes I_{7}$\\
$|geee\rangle_{1326}$ & $|g\rangle_{5}$ & $U_{4}\otimes\sigma_{7x}$
&
$|egge\rangle_{1326}$ & $|g\rangle_{5}$ & $\sigma_{4x}\otimes I_{7}$\\
$|gggg\rangle_{1326}$ & $|e\rangle_{5}$ & $\sigma_{4z}\otimes U_{7}$
&
$|ggeg\rangle_{1326}$ & $|e\rangle_{5}$ & $\sigma_{4z}\otimes\sigma_{7z}$\\
$|gggg\rangle_{1326}$ & $|g\rangle_{5}$ & $I_{4}\otimes U_{7}$ &
$|ggeg\rangle_{1326}$ & $|g\rangle_{5}$ & $I_{4}\otimes\sigma_{7z}$\\
$|eggg\rangle_{1326}$ & $|e\rangle_{5}$ & $U_{4}\otimes U_{7}$ &
$|geeg\rangle_{1326}$ & $|e\rangle_{5}$ & $\sigma_{4x}\otimes\sigma_{7z}$\\
$|eggg\rangle_{1326}$ & $|g\rangle_{5}$ & $\sigma_{4x}\otimes U_{7}$
&
$|geeg\rangle_{1326}$ & $|g\rangle_{5}$ & $U_{4}\otimes\sigma_{7z}$\\
$|ggee\rangle_{1326}$ & $|e\rangle_{5}$ &
$\sigma_{4z}\otimes\sigma_{7x}$ &
$|ggge\rangle_{1326}$ & $|e\rangle_{5}$ & $\sigma_{4z}\otimes I_{7}$\\
$|ggee\rangle_{1326}$ & $|g\rangle_{5}$ & $I_{4}\otimes\sigma_{7x}$
&
$|ggge\rangle_{1326}$ & $|g\rangle_{5}$ & $I_{4}\otimes I_{7}$\\
$|gegg\rangle_{1326}$ & $|e\rangle_{5}$ & $\sigma_{4x}\otimes U_{7}$
&
$|gege\rangle_{1326}$ & $|e\rangle_{5}$ & $\sigma_{4x}\otimes I_{7}$\\
$|gegg\rangle_{1326}$ & $|g\rangle_{5}$ & $U_{4}\otimes U_{7}$ &
$|gege\rangle_{1326}$ & $|g\rangle_{5}$ & $U_{4}\otimes
I_{7}$\\\hline
\end{tabular}\\

Finally, it is necessary to give a brief discussion on the
experimental matters. We consider the typical experimental values of
the parameters for Rydberg atoms with principal quantum numbers 49,
50, 51, the radiative time is about $T_{r}=3\times10^{-2}$ s, and
the coupling constant is $g=2\pi\times24$ kHz. For a normal cavity,
the decay time can reach $T_{c}=1.0\times10^{-3}$ s. Then we get
that the interaction time of atom and cavity is on the order of
$10^{-4}$ s. Hence, the total time for the whole system is much
shorter than $T_{r}$ and $T_{c}$, so that the present scheme might
be realizable based on cavity QED techniques.

\section{Summary}

We have proposed a simple scheme to realize the controlled
teleportation of an arbitrary two-atom state $|\phi \rangle
_{12}=a|gg\rangle _{12}+b|ge\rangle _{12}+c|eg\rangle
_{12}+d|ee\rangle _{12}$ in driven cavity QED. Two atoms exist in
single-mode cavity and are driven by a classical field. By detecting
the states of atoms, we can achieve the controlled teleportation.
The scheme is insensitive to the cavity decay and the thermal field.
Meanwhile this idea can be used to teleport an arbitrary two-atom
state from Alice to a receiver Bob via the control of n agents. The
probability of the success can reach 1.0. Two-atom maximally
entangled state can be readily prepared by atom-cavity field
interaction, which has been experimentally realized[24]. Thus this
scheme is realizable with
technique presently available. \\
\noindent\textbf{Acknowledgement}

\indent We thank Professor Guang-Can Guo for helpful suggestions.
This work is supported by the National Natural Science Foundation of
China under Grant No. 10774088,  Natural Science Foundation of Hubei
Province of China under Grant No. 2006ABA055 and the Postgraduate
Programme of Hubei Normal University under Grant No. 2007D20.


\begin{thebibliography}{100}
\bibitem{1} C. H. Bennett, et al., Phys. Rev. Lett. \textbf{70} 1895 (1993).

\bibitem{2} A. Barenco, D. Deutsch and A. Ekert, Phys. Rev. Lett.\textbf{74}
 4083(1995).

\bibitem{3} T. Sleator and H. Weinfurter, Phys. Rev. Lett. \textbf{74}
4087 (1995).

\bibitem{4} S. B. Zheng, Opt. Commun. \textbf{167} 111(1999).

\bibitem{5} S. Bandyopadhyay, Phys. Rev. A \textbf{62} 012308(2000).

\bibitem{6} D. Bouwmeester, et al., Nature (London) \textbf{390} 575(1997).

\bibitem{7} A. Furusawa, et al., Science \textbf{282} 706(1998).

\bibitem{8} M. A. Nielsen, E. Knill, and R. Laflamme, Nature (London) \textbf{
396} 52 (1998).

\bibitem{9} A. Karlsson and M. Bourennane, Phys. Rev. A \textbf{58}
4394 (1998).

\bibitem{10} O. Cohen, Phys. Rev. Lett. \textbf{80} 1121(1998).

\bibitem{11} F. L. Yan and D. Wang, Phys. Lett. A \textbf{316} 297(2003).

\bibitem{12} C. P. Yang, Shih-I Chu, and S. Y. Han, Phys. Rev. A \textbf{70}
 022329 (2004).

\bibitem{13} M. Riebe et al., Nature(London) \textbf{429} 734 (2004).

\bibitem{14} M. D. Barrett et al., Nature(London) \textbf{429} 737 (2004).

\bibitem{15} M. H. Y. Moussa, Phys. Rev. A \textbf{55} R3287 (1997).

\bibitem{16} N. G. de Almeida et al., Phys. Rev. A \textbf{62} R010101 (2000).

\bibitem{17} S. B. Zheng, Phys. Rev. A \textbf{69} 064302 (2004).

\bibitem{18} L. Ye and G. C. Guo, Phys. Rev. A \textbf{70} 054303 (2004).

\bibitem{19} L. H. Jin, X. R. Jin, and S. Zhang, Phys. Rev. A \textbf{72}
024305 (2005).

\bibitem{20} J. Lee, H. Min and S. D. Oh, Phys. Rev. A \textbf{66}
052318 (2002).

\bibitem{21} G. Rigolin, Phys. Rev. A \textbf{71}032303 (2005).

\bibitem{22} F. G. Deng et al., Phys. Rev. A \textbf{72} 022338 (2005).

\bibitem{23} S. B. Zheng, Phys. Rev. A \textbf{68} 035801 (2003).

\bibitem{24} S. Osnaghi et al., Phys. Rev. Lett. \textbf{87} 037902 (2001).
\end{thebibliography}
\end{document}